\documentclass[twocolumn,showpacs,preprintnumbers,amsmath,amssymb,prl]{revtex4}

\usepackage{graphicx}
\usepackage{dcolumn}
\usepackage{bm}

\newcommand{\bfr}{{\bf r}}

\begin{document}

\title{Angular momentum exchange between coherent light and matter fields}
\author{T.~P. Simula,$^1$ N. Nygaard,$^1$ S.~X. Hu,$^2$ L.~A. Collins,$^2$ B.~I. Schneider,$^3$ and K. M\o lmer,$^1$}
\affiliation{$^1$Lundbeck Foundation Theoretical Center for Quantum System Research, Department of Physics and Astronomy, University of Aarhus, DK-8000 Aarhus C, Denmark}
\affiliation{$^2$Theoretical Division, Los Alamos National Laboratory, Los Alamos, New Mexico 87545, USA}
\affiliation{$^3$Physics Division, National Science Foundation, Arlington, Virginia 22230 USA and Electron and Optical Physics Division, National Institute of Standards and Technology, Gaithersburg, Maryland 20899, USA}

\date{\today}

\begin{abstract}
Full, three dimensional, time-dependent simulations are presented demonstrating the quantized
transfer of angular momentum to a Bose-Einstein condensate from a laser carrying 
orbital angular momentum in a Laguerre-Gaussian mode. The process is described in terms 
of coherent Bragg scattering of atoms from a chiral optical lattice. The transfer 
efficiency and the angular momentum content of the output coupled vortex state are 
analyzed and compared with a recent experiment.  
\end{abstract}

\maketitle

Photons possess both translational and spin-angular momentum, where the latter is 
associated with different polarization states of the light wave. Light may also be 
prepared in states with well-defined orbital angular momentum as demonstrated through 
the generation of Laguerre-Gaussian (LG) beams \cite{Allen1999a, Molina2007a}. Coherent
control of the interaction between light and matter facilitates
macroscopic transfer of those quantities from one medium to another,
opening up exciting and novel possibilities for quantum information engineering such
as storing and transferring light in ensembles of atoms \cite{Dutton2004a,Ginsberg2007a}.

In a recent experiment, orbital angular momentum was controllably transferred in discrete quanta from laser photons to a
Bose-Einstein condensate (BEC) of atoms \cite{Andersen2007a}. In this
type of experiment the creation of a quantized vortex via coherent transfer
of angular momentum from photons to atoms is direct and therefore does
not rely on mechanical rotation or subsequent (thermo)dynamical
equilibration. Furthermore, this process provides a clean way of
preparing topologically-distinct quantum states and enables the
creation of genuinely multiply-charged vortices of any integer winding
number. It also serves as a prototype for coherent generation of more
complicated states by employing the spatial phase of the wavefunction.
Further development may produce a valuable tool that could be employed in
applications requiring a device capable of preparing specific quantum mechanical states 
needed for quantum information processing. In this paper, we model the full 3D dynamics
of this coherent quantum process, treating the light fields within the
semiclassical approximation.

\begin{figure}[!h] 
   \centering
   \includegraphics[width=8.6cm]{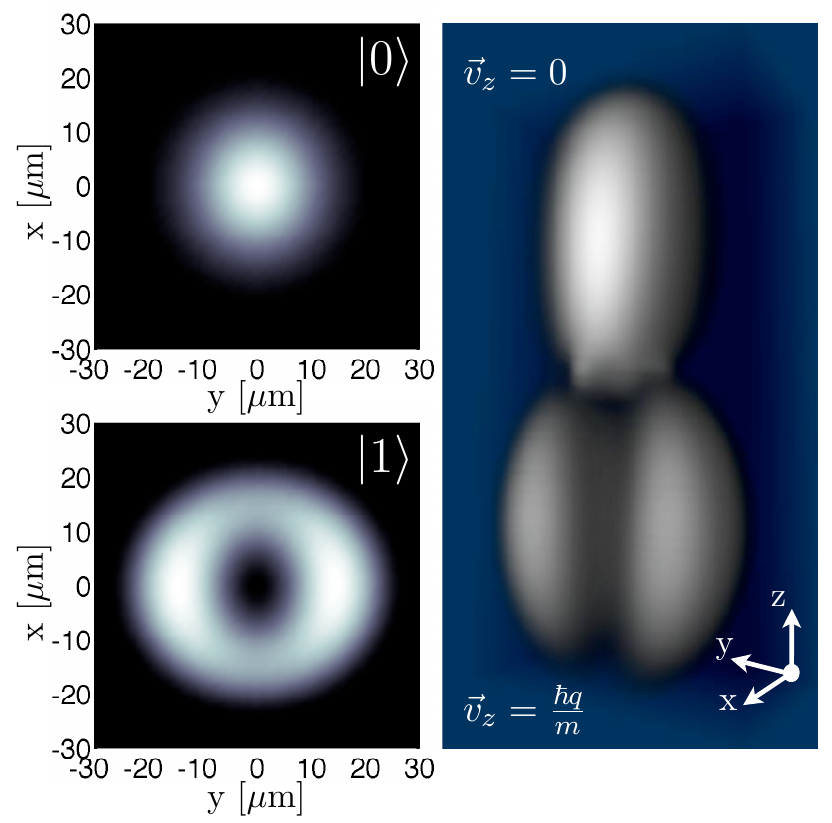} 
   \caption{(Color online) Visualization of the computed condensate density (rhs) after the Bragg scattered vortex state, denoted by $|1\rangle$, has spatially separated from the initial ground state, $|0\rangle$, atoms. The frames in the left are column densities integrated along the $z$-axis of those two distinct angular/linear momentum states. The component $|1\rangle$ is traveling to the direction of the negative $z-$ axis and the state $|0\rangle$ is at rest in the laboratory.}
   \label{fig1}
\end{figure}

Figure~\ref{fig1} illustrates the generation of a rotating cloud of atoms, which travels at a large velocity with respect to its stationary source condensate. In the experiment \cite{Andersen2007a}, this is achieved by applying to the initial BEC suitably-tuned 
counter-propagating Gaussian (G)  and Laguerre-Gaussian (LG)
laser beams. The applied light is linearly polarized and therefore has
zero-average spin-angular momentum. Linear and orbital angular momenta
carried by the photons are consequently transferred to the BEC in a
coherent two-photon Raman process in which a condensate atom absorbs
one unit, $\hbar$, of angular momentum and one unit, $\hbar k$, of
linear momentum from a LG photon. Correspondingly, in order to
satisfy energy and momentum conservation laws, the atom emits a
stimulated photon to the G field acquiring a further $\hbar k$ of linear
momentum from the recoiling photon. The linear momentum of each G photon is set approximately equal to the respective LG photon. Alternatively, the process
described above may be cast in terms of coherent Bragg scattering of
atoms from a traveling optical diffraction
grating \cite{Kozuma1999a,Blakie2002a}. In this picture, the combined
effect of the counter-propagating laser fields on the BEC may be
accounted for in the electric-dipole approximation by a chiral light-shift
potential:
\begin{equation}
V_{\rm Bragg}(\bfr,t)=  |A_{\rm G}|^2 + |A_{\rm LG}|^2 + 2A_{\rm G}^*A_{\rm LG}\cos(qz+\Delta\omega t+\phi),
\label{eq1}
\end{equation}
where $q=2k$, $\Delta\omega$ is the frequency difference between the G and LG beams, and  $\phi(x,y)$ is the spatial phase of the LG beam corresponding to a $2\pi$ phase winding around the $z-$axis.  The radial dependence of the laser fields in the $x-y$ plane ($r^2=x^2+y^2$) in terms of the intensities, $A_0$ and $A_1$, and the beam waists, $\sigma_{\rm G}$ and $\sigma_{\rm LG}$, is
\begin{equation}
A_{\rm G} =\sqrt{A_0}e^{-r^2/\sigma_{\rm G}^2} 
\textrm { and  }
A_{\rm LG}=\sqrt{A_1} r e^{-r^2/\sigma_{\rm LG}^2}. 
\end{equation}
The resulting mean-field Hamiltonian operator
\begin{equation}
\hat H=-\frac{\hbar^2}{2m}\nabla^2  +  g|\Psi(\bfr,t)|^2 +  V_{\rm trap}(\bfr) +V_{\rm Bragg}(\bfr,t) 
\end{equation}
has the Gross-Pitaevskii form containing the time-dependent potential, $V_{\rm Bragg}$, which describes the coherent atom-photon coupling, and an external harmonic trap potential
\begin{equation}
V_{\rm trap}(\bfr) = \frac{m}{2}\left(   \omega_x^2x^2  + \omega_y^2y^2 +  \omega_z^2z^2 \right) .
\end{equation}

The main computational difficulty in the full numerical treatment of a typical experiment of 
Bragg scattering of a BEC is caused by the high-frequency oscillation of the optical diffraction 
potential, requiring an extremely fine spatial grid and a small time step in the 
temporal propagation of the wavefunction. The formal solution for this 
Hamiltonian system for short time intervals, $\delta t$,    
\begin{equation}
\Psi(\bfr,t+\delta t)=e^{-i\hat H(t) \delta t/\hbar}\Psi(\bfr,t)
\end{equation}
involves an exponentiation of a sum of non-commuting operators. To
tackle these issues, we have developed an efficient, scalable parallel code for the
solution of the time-dependent non-linear Schr\"odinger equation. The method of choice 
combines a finite-element discrete variable spatial representation (FEDVR) with a fourth 
order split-operator technique for the time evolution
\cite{Schneider2005a}. The power of this procedure derives from the sparse
representation of the Hamiltonian matrix in combination with an
underlying polynomial basis that enables an efficient computation of the
matrix exponentials from a quadrature formula. In practice, the
representation of the generalized non-linear and time-dependent
potential is diagonal, while the kinetic energy part in each spatial
dimension is sparse and may be split into a sum of two block diagonal
matrices. Finally, the code is parallelized applying a
2D cartesian communicator virtual topology within the message passing
interface (MPI) protocol, resulting in marked savings in both memory
and computational time.

In accordance with the experiment \cite{Andersen2007a}, our system consists of $N=2\times 10^6$ interacting $^{23}$Na atoms placed in an anisotropic trap with harmonic frequencies,  $\{\omega_x,\omega_y,\omega_z\}=40\pi \times \{2,\sqrt{2},1\}$Hz. The coupling constant, $g=4\pi\hbar^2 a/m$, is expressed in terms of the $s-$wave scattering length, $a=2.75$nm, and the mass, $m$, of an atom. The laser field intensities are related by $A_1=4.3\times10^{-3} A_0$, their waists are, $\sigma_{\rm G}=175 \mu$m and $\sigma_{\rm LG}=85 \mu$m, and their temporal shape is modeled by a square pulse of varying duration. The frequency difference between the G and LG fields is
\begin{equation}
\Delta\omega =\Delta\omega_0  + \delta\omega
\end{equation}
where $\Delta\omega_0 = \hbar q^2/2m = 5000\;\omega_z$ and $\delta\omega$ is the detuning from the first order linear Bragg scattering resonance. Since the  chemical potential, $\mu/\hbar \approx 38 \; \omega_z \ll \Delta \omega_0 $, the mean-field shift, $\mathcal{O}(\mu)$, to the resonance frequency is almost negligible.

\begin{figure}[!t] 
   \centering
   \includegraphics[width=8.6cm]{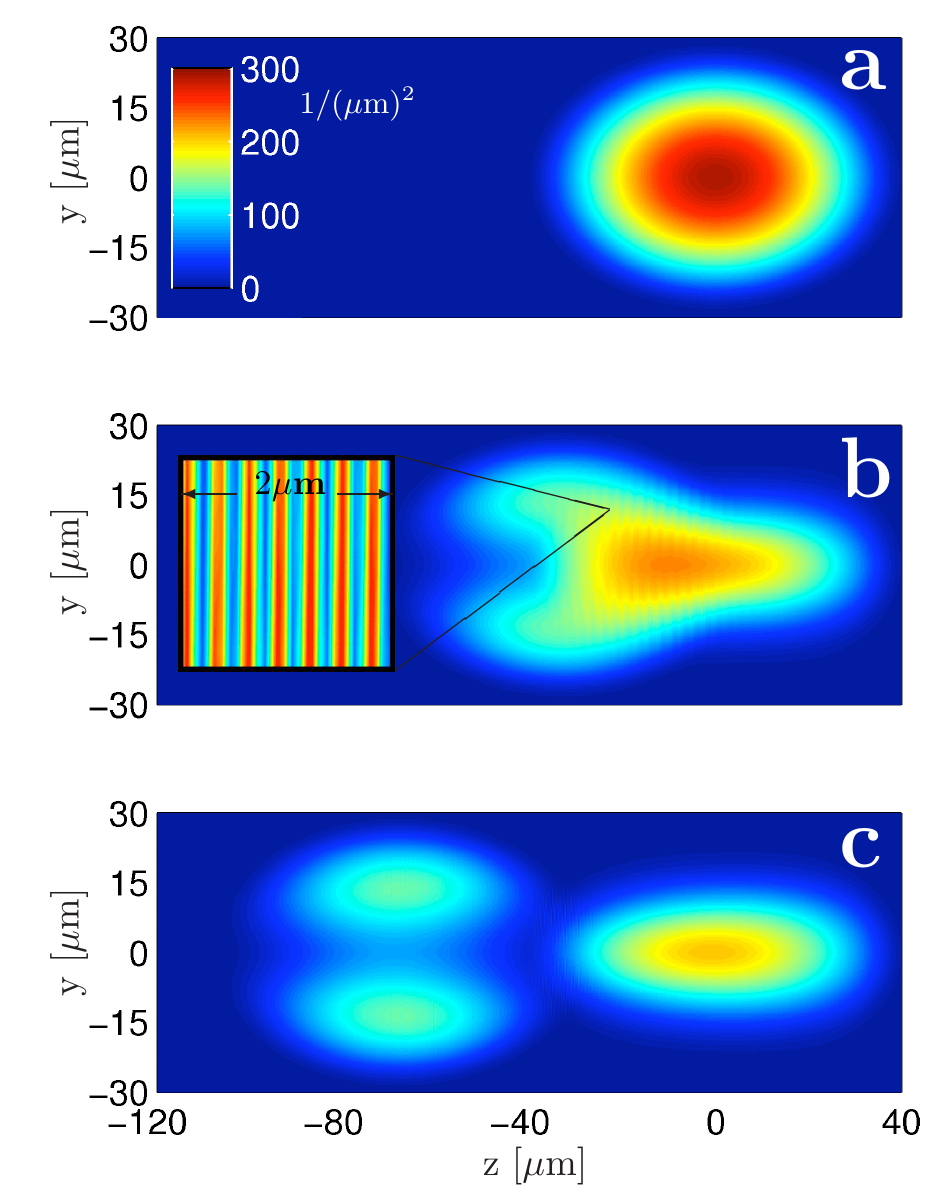} 
   \caption{(Color online) Snapshots of the system at three different times, $t =$ (0, 0.6, and 1.2) ms (a-c). Frames show column densities, $\int|\phi(\bfr)|^2 \;{\rm d}x$.  Part of the frame (b) has been magnified in the region $\{z,y\}=\{[-24,-22],[10,12]\} \;\mu$m shown in the inset to resolve the high frequency interference structure. Bragg pulse duration, $\tau = 77 \mu $s, $A_0=500\; \hbar\omega_z$ and $\delta w = 0\; \omega_z$.}
   \label{fig2}
\end{figure}

Figure \ref{fig2} displays some typical numerical results showing side-view column densities of the system at three different times. Before the laser fields are applied, the condensate is in the interacting ground state of the harmonic trap, $|0\rangle$, and the spatial density of particles is smooth as shown in Fig.~\ref{fig2}(a). As soon as the chiral optical diffraction grating is initiated, part of the original condensate scatters into a traveling vortex state, $|1\rangle$. The traveling $|1\rangle$ and stationary $|0\rangle$ components interfere in their overlap region, producing a fine helix of fringes, not visible in the main frame of Fig.~\ref{fig2}(b). The fringe spacing is determined by the relative velocity, $v=\hbar q/m$, between the two components. The wavelength of these density modulations, magnified in the inset of Fig.~\ref{fig2}(b), falls beyond current experimental imaging resolution. Finally, Fig.~\ref{fig2}(c) displays the system shortly after the two components $|1\rangle$ and $|0\rangle$ have spatially separated. The $|0\rangle$ state has become slightly tapered towards the direction of the motion of the $|1\rangle$ component due to the 
interactions rendering the radial expansion $z-$dependent as shown in Fig.~\ref{fig2}(c). At the end of the Bragg pulse, the trap is also switched off, allowing the system to expand ballistically. In Fig.~\ref{fig2} the pulse duration is chosen so that the resulting two momentum components have equal populations, thus implementing an atomic beam splitter. 

\begin{figure}[t] 
   \centering
   \includegraphics[width=8.6cm]{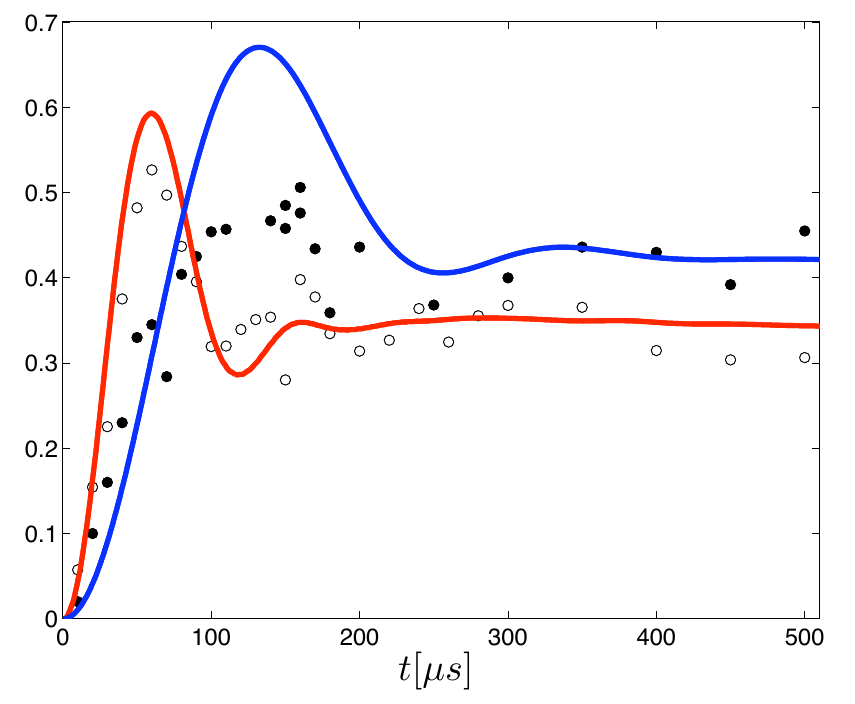} 
   \caption{(Color online) Fraction of particles in the vortex state $|1 \rangle$ as a function of the pulse duration. Experimental data sets obtained for Bragg frequencies 97.5 kHz ($\bullet$) and 100 kHz ($\circ$) are obtained by separately measuring particle number in both states $|0 \rangle$ and $|1 \rangle$ after TOF \cite{NISTdata}. Computed curves show the expectation values, $\langle L_z\rangle/N$, for different detunings, $\delta\omega=50\; \omega_z$ (blue) and $\delta \omega=175\; \omega_z$ (red), and the respective intensities $A_0=460\; \hbar\omega_z$ and $A_0=950\; \hbar\omega_z$. }
   \label{fig3}
\end{figure}

In Fig.~\ref{fig3}, we display the fraction of particles in state $|1 \rangle$ as functions of pulse duration. The experimental data for two different Bragg frequencies, 97.5 kHz ($\bullet$) and 100 kHz ($\circ$), are shown \cite{NISTdata}. Due to a systematic centre-of-mass motion of the condensate, these frequencies are Doppler shifted from the center of the Bragg resonance. Each point in the experimental data correspond to a separate experiment and they are obtained by explicitly measuring the populations in states $|0 \rangle$ and $|1 \rangle$ after they have spatially separated during a few milliseconds of time-of-flight (TOF). To compare the numerical solution with the experimental data, we have plotted the expectation value of the $z-$component of the total angular momentum per particle, $\langle L_z \rangle / N$, as a function of the pulse duration shown in Fig.~\ref{fig3} for two different pulse parameters $A_0=460\; \hbar\omega_z,\delta w=50\;\omega_z$ (blue) and $A_0=950\; \hbar\omega_z, \delta w = 175\; \omega_z$ (red). Since each atom in the traveling vortex state carries a single unit of angular momentum, the computed curves may be viewed as the fraction of the transferred population to the vortex state $|1\rangle$ as a function of pulse duration. We have also computed the expectation value of the total linear momentum per particle, $\langle P_z \rangle / N$, scaled by $q$. Such curves are indistinguishable from the ones shown, leading to the conclusion that the transfer of linear and angular momenta is fully correlated, verifying the quantized transfer mechanism between the light and matter fields. During its exposure to the light field, the condensate executes Rabi cycling between the ground state $|0\rangle$ and the vortex state $|1 \rangle$ due to the photon-mediated coupling between them, as is evident in Fig.~\ref{fig3}. The maximum atom transfer is limited by the overlap between the ground state condensate and the spatial shape of the optical Bragg grating. The discrepancy between the experimental data and computed transfer curves may be accounted for by experimental uncertainties, such as the residual, in-trap, center-of-mass motion of the condensate.

\begin{figure}[!t] 
   \centering
   \includegraphics[width=8.6cm]{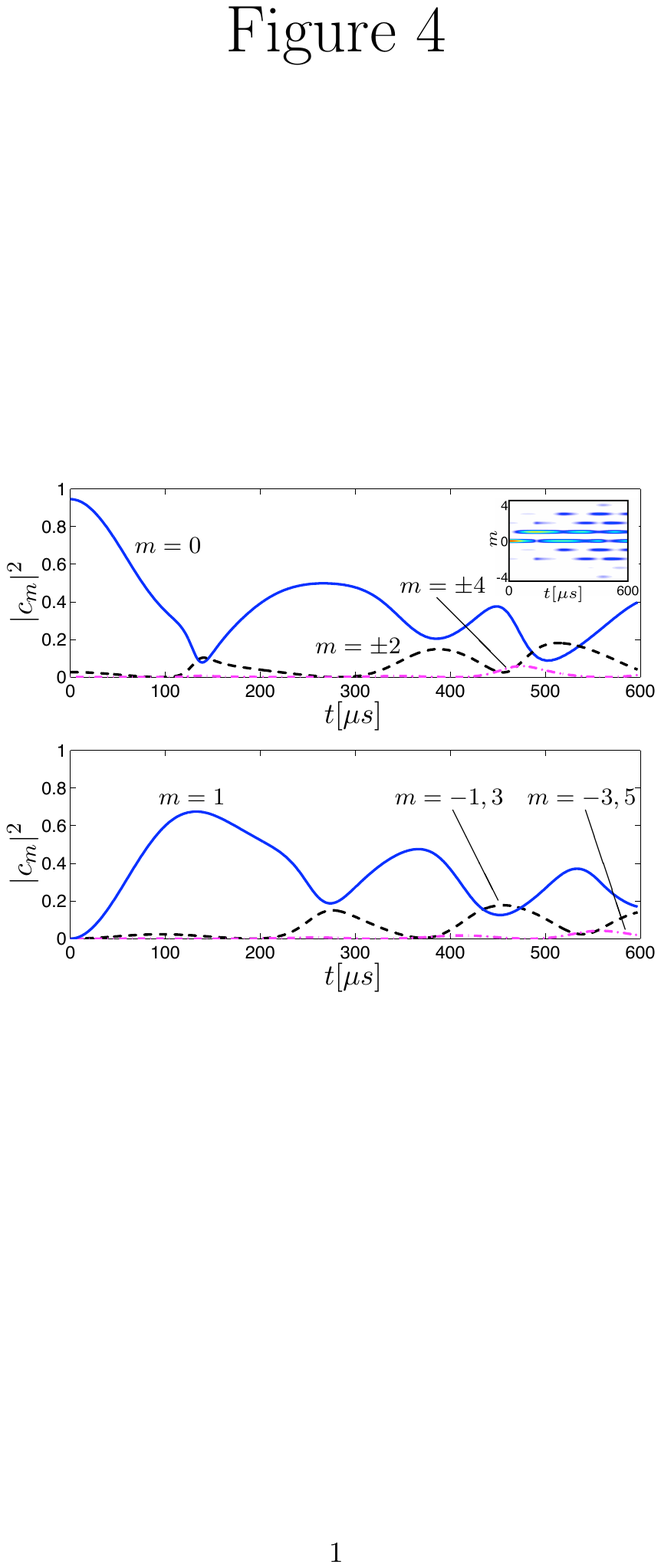} 
   \caption{(Color online) Populations of different angular momentum states as functions of pulse duration. Upper panel contains the even parity, $p=0$, states and the lower panel is for odd parity, $p=\hbar q$, states. Different $m$ values are indicated in the picture. The subfigure in top frame shows evolution of populations of all $m$ states in a single plot as function of pulse duration. The pulse parameters are as in Fig.~\ref{fig2}. }
   \label{fig4}
\end{figure}

Since the trap is fully anisotropic, the condensate ground state is not an angular momentum eigenstate but is instead described in terms of a superposition of different angular momentum basis states. In order to analyze the content and transfer of angular momentum to the condensate, we express its wavefunction, $\Psi(\theta,z)=\sum_{p,m}c_{p,m} \exp (ipz/\hbar  + im\theta)$, in terms of the different $m$ states. Since the momentum spread around the stationary and traveling momentum components is narrow, we only consider two discrete linear momentum values $p=0$ and $p=\hbar q$. In Fig.~\ref{fig4}, we plot the normalized probability densities $|c_{p,m}|^2$ of the projection of the wavefunction to different angular momentum states as functions of pulse duration. In the top frame $p=0$, and only even $m$ values are occupied. Similarly, in the lower frame $p=\hbar q$, and only odd $m$ states are populated. The moving and stationary components have odd and even parities in the $x-y$-plane, respectively. As is seen in Fig.~\ref{fig4}(a), the initial ground state already contains a 3\% admixture of $m=\pm 2$ states due to the trap asymmetry. In the course of time the odd and even $m$ states become populated symmetrically around the respective $m=1$ and $m=0$ states. Coupling with the laser fields induces the dominant dipole transitions between $m=0$ and $m=1$ states and the undulations in the side band $m$ states arise due to the mean-field interactions, as is verified by a comparative simulation in which the interactions are neglected. In the inset of Fig.~\ref{fig4} we have plotted populations of all \{$p,m$\} states as functions of time showing the population spreading to larger $m$ values. 

In conclusion, by modeling the recent experiment, we have numerically demonstrated coherent transfer of orbital angular momentum from Laguerre-Gaussian photons to Bose-Einstein condensed atoms in agreement with the experimental observations. We have employed a powerful numerical method, which enables us to calculate the full 3D dynamics of this system. Altering the squeezing of the trap frequencies permits the manipulation of the relative populations of different angular momentum states. This method may be used to implement a beam splitter into two topological states displaying the potential for generation of robust phase qubits. Generalizing, the LG states of light form a complete orthonormal basis set which may be deployed to gain an access to an infinite-dimensional Hilbert space of quantum states for quantum information processing. The underlying principle allows for the creation of exotic optical lattice configurations and the study of persistent currents in toroidal trap configurations \cite{Ryu2007a}. The latter is a promising starting point to investigate the feasibility of creating macroscopically entangled superflow states of opposite circulation.

\begin{acknowledgments}
We are grateful to the Authors of Ref.~\cite{Andersen2007a} for insightful discussions and for the permission to include the original data from their experiment in this paper. Allan Rasmusson and Morten Lervig at The Centre for Advanced Visualization and Interaction (CAVI) are acknowledged for their assistance in visualizing our numerical data.

\end{acknowledgments}


\begin{thebibliography}{90}
\bibitem{Allen1999a}
L. Allen, S.~M. Barnett and M.~J. Padgett, \emph{Optical Angular Momentum} (IOP Publishing, Bristol, 2003). 
\bibitem{Molina2007a}
G. Molina-Terriza, J.~P. Torres and L. Torner, \emph{Nat. Phys.} {\bf 3}, 305 (2007). 
\bibitem{Dutton2004a}
Z. Dutton and J. Ruostekoski, \emph{Phys. Rev. Lett.} {\bf 93}, 193602 (2004).
\bibitem{Ginsberg2007a}
N.~S. Ginsberg, S.~R. Garner, and L. Vestergaard Hau, \emph{Nature} {\bf 445}, 623 (2007).
\bibitem{Andersen2007a}
M.~F. Andersen, C. Ryu, P. Clad\'e, V. Natarajan, A. Vaziri, K. Helmerson and W.~D. Phillips, \emph{Phys. Rev. Lett.} {\bf 97}, 170406 (2006).
\bibitem{Kozuma1999a}
M. Kozuma, L. Deng, E. W. Hagley, J. Wen, R. Lutwak, K. Helmerson, S.~L. Rolston and W.~D. Phillips, \emph{Phys. Rev. Lett.} {\bf 82}, 871 (1999).
\bibitem{Blakie2002a}
P.~B. Blakie, R.~J. Ballagh and C.~W. Gardiner, \emph{Phys. Rev.  A.} {\bf 65}, 033602 (2002).
\bibitem{Schneider2005a}
B.~I. Schneider, L.~A. Collins and S.~X. Hu, \emph{Phys. Rev. E} {\bf 73}, 036708 (2006).
\bibitem{NISTdata}
NIST Laser Cooling and Trapping Group, private communication.
\bibitem{Ryu2007a}
C. Ryu, M.~F. Andersen, P. Clad\'e, V. Natarajan, K. Helmerson and W.~D. Phillips, (to be published).
\end{thebibliography}
\end{document}